\documentclass[superscriptaddress,aps,pra,twocolumn,showpacs,nofootinbib,longbibliography]{revtex4-1}
\usepackage{algorithm}
\usepackage{algpseudocode,caption}
\usepackage{amsmath,amsthm}
\usepackage{braket}
\usepackage{amsmath}
\usepackage{array}
\usepackage{amsfonts}
\usepackage{amssymb}\usepackage[T1]{fontenc}
\usepackage{latexsym}
\usepackage{amssymb}
\usepackage[colorlinks=true,citecolor=blue,urlcolor=blue]{hyperref}
\usepackage{color}
\usepackage{graphics,epstopdf}
\usepackage{soul}
\usepackage[demo]{graphicx}
\usepackage{subcaption}
\usepackage{lipsum}
\usepackage{float}
\usepackage{adjustbox}
\usepackage[normalem]{ulem}
\usepackage{listings}

\newcommand{\be}{\begin{equation}}
\newcommand{\ee}{\end{equation}}
\newcommand{\ba}{\begin{eqnarray}}
\newcommand{\ea}{\end{eqnarray}}

\begin{document}
	
\title{A universal whitening algorithm for commercial random number generators}

\author{Avval Amil}
\email{avval2801@gmail.com}
\affiliation{Department of Computer Science and Engineering, I.I.T. Delhi, Hauz Khas, New Delhi - 110016, India}

\author{Shashank Gupta}
\email{shashank@qnulabs.com}
\affiliation{QuNu Labs Pvt. Ltd., M. G. Road, Bengaluru, Karnataka 560025, India}

\date{\today}
\begin{abstract}

Random number generators are imperfect due to manufacturing bias and technological imperfections. These imperfections are removed using post-processing algorithms that in general compress the data and do not work in every scenario. In this work, we present a universal whitening algorithm using n-qubit permutation matrices to remove the imperfections in commercial random number generators without compression. Specifically, we demonstrate the efficacy of our algorithm in several categories of random number generators and its comparison with cryptographic hash functions and block ciphers. We have achieved improvement in almost every randomness parameter evaluated using ENT randomness test suite. The modified random number files obtained after the application of our algorithm in the raw random data file pass the NIST SP 800-22
tests in both the cases: 1. The raw file does not pass all
the tests. 2. The raw file also passes all the tests.

\end{abstract}

\maketitle
\section{Introduction}
Whitening algorithms have demonstrated remarkable improvement in the quality of all kinds of random number generators \cite{Cusik09, Thamrin09, Sam10}. It was shown that the raw random numbers fail some of the statistical randomness tests that pass after the application of whitening algorithms. In traditional whitening algorithms, linear combination of the raw random number sequences with the benchmark random sequence resulted in increased variance of the sequence and thus the quality of the randomness \cite{Sam10}. The sole objective of a whitening algorithm is to improve the quality of the random number generators.

Random number generators can be categorised depending upon the notion of the randomness \cite{Bera17, Vaisakh22}: 1. Deterministic or pseudo random number generators (PRNGs). 2. Epistemic or classical true random number generators (TRNGs). 3. Ontic or quantum random number generators (QRNGs). Recently a general trend towards QRNGs can be seen in commercial random number market \cite{IDQ17, Tropos, Quintessence20, Toshiba} because of their potential to harness random numbers from rather simple and intrinsic random quantum phenomenon as compared to complex chaotic classical phenomenon in classical TRNGs \cite{Toni_1, Toni_2} and deterministic PRNGs \cite{James90} that are prone to entropic starvation in real world applications \cite{Peter21}. That said, all three category of random number generators are reported as imperfect due to manufacturing bias or technological imperfections \cite{Hurley17}.

Among random number generators, PRNGs are most commonly used (e.g. C++ rand() function, Linux/dev/urandom (random)). Several other PRNGs have been proposed for resource contrained IoT devices \cite{Orue17}. Baldanzi et al. recently presented a
cryptographically secure deterministic random bit generators (DRBGs) by analyzing different cryptographic
algorithms, such as SHA2, AES-256 CTR mode, and triple
DES \cite{Baldanzi20}. In 2020, James et al. reviewed high quality PRNGs on the basis of Kolmogorov–Anosov theory
of mixing in classical mechanical systems \cite{James20}. Xorshift is a special PRNG under non-cryptographically secure random number generators \cite{George03, Vigna14}. The biggest issue with the PRNGs is the limited entropy injected through seeds resulting in the situation entropic starvation in the real world applications \cite{Peter21}.

The low entropy problem was addressed using hardware based classical TRNGs that were shortly substituted by QRNGs because of the simplistic phenomenon and ontic nature of the randomness \cite{Nie16, Ma16, Tobias21}. Remarkable progress has happended over the years in terms of speed, size, cost, self testable in the field of QRNGs. Several commercial QRNGs are available in the market in several form factors like PCI card based from IDQ and Quintessance labs, USB, chip etc. Commercial QRNGs require postprocessing using some whitening algorithms to remove manufacturing biases in the outputs \cite{Hurley17}. 

There are several postprocessing algorithms developed for improving specific random number generators \cite{Hayashi16, Huang18, Tang19}. Our motivation is to devise a universal algorithm that can work with every random number generator. Our algorithm utilises the huge information space associated with the n-qubit permutation matrices to remove the bias or higher order correlations among the generated random number sequence. Permutation matrices have been efficient in expanding entropy, maintaining Shannon perfect secrecy and fundamental in laying the foundation of the quantum safe cryptography \cite{Randy20, Kuang21, Randy21, Lou21}. Permutation matrices can be realised using traditional classical as well as quantum computing systems \cite{Shende03}. Hence, any algorithmic development with permutation matrices offer huge agility in the current as well as future cryptographic infrastructure.

In the present work, we demonstrate the action of an n-qubit permutation matrices based whitening algorithm in removing bias and other defects in commercial PRNGs, classical TRNGs and QRNGs. In particular, we analyse the efficacy of the whitening algorithm in improving randomness parameters of the popular PRNGs (/dev/urandom, /dev/random, C++ rand() function), ring oscillator based classical TRNGs and commercial QRNGs, and compare it with cryptographic hash functions (SHA-256), block ciphers (DES-CBC, AES-256-CTR).
We demonstrate that our algorithm works in these scenarios without compressing the random data.   

The paper is organised in the following way. In Sec.(\ref{Prelim}), we recapitulate the set of n-qubit permutation matrices and the generation of an n-qubit permutation matrices using QRNG data from the Qosmos (QNu Labs Entropy-as-a-Service). In Sec.(\ref{WA}), we discuss the universal whitening algorithm using the generated permutation matrices. In Sec. (\ref{TR}), we present our results in the following scenarios, A. PRNGs. B. Comparison with Cryptographic Hash functions C. Comparison with Block cipher DES-CBC and AES-256-CTR. D. TRNG. E. QRNGs. Finally, we 
conclude in Sec.(\ref{Conclusion}) with some future offshoots of the present analysis.

\section{Preliminaries}
\label{Prelim}
Some definitions and software we have used to test the data created by our algorithm are described now - \\ \\
\textit{Whitening Algorithms:} A whitening algorithm is a technique that is applied on a stream of random numbers (entropy stream) which reduces the bias and correlation for the data (can be in the form of bytes or individual bits or both) and improves the random characteristics for the stream. In many of the commonly used whitening algorithms, multiple input bits are used to create a single output bit, and thus the size of the data reduces. As we will see in the following section, this is in contrast with our whitening algorithm which preserves the size of the data. Some famous whitening algorithms are-
\begin{itemize}
    \item XOR : In this technique, two streams of random numbers are used to create a single entropy stream. One bit is taken from each stream and their exclusive OR is taken to give the output bit. Two bits are used up to create one bit here. 
    \item Cryptographic Hash Functions : Various cryptographic hash functions exist which also improve the randomness of the entropy stream. These hash functions (e.g. SHA-256, SHA-512 etc.) reduce the size of the data.
    \item Von Neumann's Technique: This is a simple technique in which bits of the input stream are considered two at a time, if the two bits are the same (either `00' or `11') they are discarded, if the bits are `01' or `10, then the output bit is `0' or `1' respectively. It can be seen that this technique uses more than two bits on average to generate a single bit of the output.
\end{itemize}
\textit{ENT Randomness Test:} This is program used to test the randomness of a random number sequence \cite{ENT}. It can be used in two modes - the bit mode which works by considering in chunks of single bits, or the byte mode which considers data to be chunks of 8 bits at a time. For all our testing, we have worked with the byte mode. This test gives results for five parameters related to random data on the basis of which the quality of the random numbers being tested can be gauged. These five parameters are - entropy, chi square distribution, arithmetic mean, Monte Carlo approximation of $\pi$ and serial correlation coefficient. By comparing the values of these parameters with their expected value for truly random data, the quality of the random number generator is judged. Note that these tests are statistical in nature and hence give more accurate results for larger sizes of the input stream. Keeping this in mind, all the data we have tested on was large in size (100MB - 1GB of random numbers).\\ \\
\textit{NIST Statistical Test Suite:} This is set of 15 tests (and various sub-tests) \cite{NIST} developed by the National Institute of Standards and Technology, USA. It was developed after DES was cracked to choose today's AES. \cite{NISTReview}. It provides a comprehensive set of tests for which a P-value is output. Higher P-values correspond to better quality random numbers (a P-value of 0 corresponds to perfectly non-random numbers while a P-value of 1 corresponds to perfect randomness) and high quality random numbers should be able to pass all of these tests.

\section{Whitening Algorithm}\label{WA}
We have used the entropy expansion algorithm \cite{Avval22} for the purpose of whitening random data from various sources. 
This algorithm breaks the input data (which here is the bit-string corresponding to the binary content of the files under consideration) into chunks and then multiplies the chunk with a randomly selected permutation matrix (a matrix in which each row and each column has exactly one non-zero entry equal to one) to create the output chunk. For example, if the input chunk is `0001' which is to be multiplied with the permutation matrix shown, below, the output chunk will be `0010'.
\begin{center}
        $\begin{bmatrix}
            1 & 0 & 0 & 0\\
            0 & 0 & 1 & 0\\
            0 & 0 & 0 & 1\\
            0 & 1 & 0 & 0
        \end{bmatrix}
        \times
        \begin{bmatrix}
            0\\
            0\\
            0\\
            1
        \end{bmatrix}
        =
        \begin{bmatrix}
            0\\
            0\\
            1\\
            0
        \end{bmatrix}$ 
\end{center}
 The output chunks are then concatenated to form another bit-string which is the binary content of the output file. For the generation of permutation matrices, we have used the Fisher-Yates shuffle algorithm, which is now given in the form of pseudocode.\\
 Note that the generation of the matrices has a random step to it ('RandomInt(1,N)' in the pseudocode) which is one reason why this whitening algorithm is not deterministic.
\begin{algorithmic}[1]
\Require RandomInt(1,$N$) returns a random integer between 1 and $N$ (inclusive) when called; Swap($a$,$b$) swaps the values of $a$ and $b$
\Ensure $P$ is a random permutation matrix
\State $i \leftarrow 1$
\While{$i \leq N$}
\State $K[i] \leftarrow$ RandomInt(1,$N$)
  \State $S[i] \leftarrow i$
  \State $j \leftarrow 1$
    \While{$j \leq N$}
        \State $P[i][j] \leftarrow 0$
        \State $j \leftarrow (j+1)$
    \EndWhile
    \State $i \leftarrow (i+1)$
\EndWhile
\State $i\leftarrow N$
\While{$i>0$}
    \State $p \leftarrow K[i]$
    \State $Swap(S[p],S[i])$
    \State $i \leftarrow (i-1)$
\EndWhile
\State $i \leftarrow 1$
\While{$i\leq N$}
    \State $P[i][S[i]] \leftarrow 1$
    \State $i \leftarrow (i+1)$
\EndWhile
\end{algorithmic}

As the size of the permutation matrix (and hence the chunks) increases, the permutation space associated with the set of the permutation matrices widens, and this leads to higher increase in randomness of the data on applying permutations. Using a greater number of permutation matrices in our set also increases the entropy.\\ Using entropy expansion as a whitening technique has a few differences from the other whitening algorithms described before - for one, (if the size of the entropy stream is a multiple of the dimension of the permutation matrix) then the size of the output is exactly the same as the size of the input; also, the output of entropy expansion on the input data may be different each time it is applied due to the random selection of the permutation matrices. (as opposed to most whitening algorithms which are deterministic in nature). 
\begin{figure}[h!]
    \centering
    \resizebox{8.4cm}{10cm}{\includegraphics{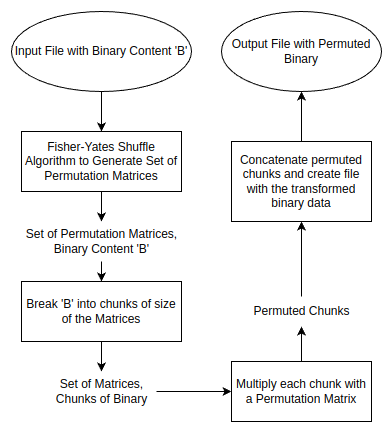}}
    \caption{Flowchart Explaining the Entropy Expansion Whitening Algorithm}
    \label{expansionalgo}
\end{figure}

\section{Testing of the Whitening Algorithm and Results}\label{TR}
We tested the whitening algorithms on several random data generated by various random number generators, and also compared the results of the whitening algorithm against commonly used cryptographic hash functions like SHA-256 as well as block ciphers like AES-256 and DES. The tests we conducted are now listed - 
\subsection{Pseudo-Random Number Generators (PRNGs)}
\subsubsection{C++ rand() Function} We generated 1GB of random numbers using the inbuilt 'rand()' function in C++. The seed taken for the generation of these numbers was decided by using 'time.h' and the current time in seconds was used as the seed. Earlier, the algorithm used by this function was a simple multiply and shift algorithm, but now it has been replaced by a more cryptographically secure algorithm. The random numbers generated passed all the NIST tests (Refer \ref{rand} - note that failures in the NIST reports are indicated by an asterisk next to the test parameters). After modification by our whitening algorithm, it was found that the modified numbers still passed all the NIST tests (Refer \ref{randmod}). The ENT test results are tabulated (Table \ref{table1}). It can be seen that except the chi-square value, all parameters show an improvement. 
\subsubsection{Linux /dev/urandom} In Unix-like operating systems, /dev/urandom is a special file that can create pseudo-random numbers using data from environmental noise as a seed. 1GB of random numbers was generated using /dev/urandom and the testing was performed on both the original numbers and the modified numbers. We saw that 2 NIST tests failed for the original random numbers (which was to be expected as /dev/urandom is known for not being cryptographically secure) (Refer \ref{urandom}), while all the tests passed for the modified numbers (Refer \ref{urandommod}). The modification was done using 32 matrices of $8192\times 8192$ size. The ENT test results are tabulated (Table \ref{table1}). Except the serial correlation coefficient, all parameters show an improvement.
\subsubsection{Linux /dev/random} Like /dev/urandom, /dev/random also creates pseudo-random numbers using data from the environment as seed, but if there is less entropy, it blocks the random numbers. This results in better quality random numbers and it was observed that all NIST tests passed for /dev/random (Refer \ref{random}). After modification, this was still the case and all NIST tests passed for the modified numbers (Refer \ref{randommod}). The ENT test results are tabulated (Table \ref{table1}). Here too, except the serial correlation coefficient, all parameters show an improvement.
\begin{widetext}
\begin{center}
\begin{table}[h!]
\centering
    \begin{tabular}{| m{2cm} | m{2cm} | m{2cm} | m{2cm} | m{2cm} | m{2cm} | m{2cm} | m{2cm} |}
        \hline
        Parameter & Input File 1& Output File 1& Input File 2& Output File 2& Input File 3& Output File 3&Ideal Value\\
        \hline \hline
        Entropy & 8.000000 & 8.000000 & 8.000000 & 8.000000 & 8.000000 & 8.000000 &8.000000\\
        \hline
        Chi-Square Distribution & 256.99 & 271.33 & 281.74 & 254.60 & 298.63 & 239.96 &256.00 \\
        \hline
        Arithmetic Mean & 127.5024 & 127.5002 & 127.5035 & 127.4985 & 127.4964 & 127.5008 &127.5\\
        \hline
        Monte Carlo value of Pi & 3.141405155 & 3.141684931 & 3.141483117 & 3.141554084 & 3.141550664 & 3.141517494 &3.141592653\\
        \hline
        Serial Correlation Coefficient & -0.000024 & 0.000022 & -0.000015 & 0.000019 & -0.000022 & 0.000032 &0.0\\
        \hline
    \end{tabular}
    \caption{Results for C++ rand() (File 1), /dev/urandom (File 2) and /dev/random (File 3) 1GB random numbers}
    \label{table1}
\end{table}
\end{center}
\begin{center}
\begin{table}[h!]
\centering
    \begin{tabular}{| m{2cm} | m{2cm} | m{2cm} | m{2cm} | m{2cm} | m{2cm} | m{2cm} | m{2cm} |}
        \hline
        Parameter & SHA-256 on C++ rand()& Modified C++ rand()& SHA-256 on /dev/urandom& Modified /dev/urandom& SHA-256 on /dev/random& Modified /dev/random&Ideal Value\\
        \hline \hline
        Entropy & 7.999999 & 8.000000 & 7.999999 & 8.000000 & 7.999999 & 8.000000 &8.000000\\
        \hline
        Chi-Square Distribution & 336.31 & 271.33 & 268.95 & 254.60 & 267.53 & 239.96 &256.00 \\
        \hline
        Arithmetic Mean & 127.4986 & 127.5002 & 127.4912 & 127.4985 & 127.5020 & 127.5008 &127.5\\
        \hline
        Monte Carlo value of Pi & 3.141316661 & 3.141684931 & 3.141813093 & 3.141554084 & 3.141714298 & 3.141517494 &3.141592653\\
        \hline
        Serial Correlation Coefficient & -0.000036 & 0.000022 & 0.000043 & 0.000019 & 0.000004 & 0.000032 &0.0\\
        \hline
    \end{tabular}
    \caption{Comparison of SHA-256 with Entropy Expansion for C++ rand(), /dev/urandom and /dev/random 1GB random numbers}
    \label{table2}
\end{table}
\end{center}
\begin{center}
\begin{table}[h!]
\centering
    \begin{tabular}{| m{2cm} | m{2cm} | m{2cm} | m{2cm} | m{2cm} | m{2cm} | m{2cm} | m{2cm} |}
        \hline
        Parameter & DES-CBC on C++ rand()& Modified C++ rand()& DES-CBC on /dev/urandom& Modified /dev/urandom& DES-CBC on /dev/random& Modified /dev/random&Ideal Value\\
        \hline \hline
        Entropy & 8.000000 & 8.000000 & 8.000000 & 8.000000 & 8.000000 & 8.000000 &8.000000\\
        \hline
        Chi-Square Distribution & 256.65 & 271.33 & 235.62 & 254.60 & 227.95 & 239.96 &256.00 \\
        \hline
        Arithmetic Mean & 127.5003 & 127.5002 & 127.5003 & 127.4985 & 127.5029 & 127.5008 &127.5\\
        \hline
        Monte Carlo value of Pi & 3.141422988 & 3.141684931 & 3.141670578 & 3.141554084 & 3.141474486 & 3.141517494 &3.141592653\\
        \hline
        Serial Correlation Coefficient & -0.000031 & 0.000022 & 0.000024 & 0.000019 & 0.000024 & 0.000032 &0.0\\
        \hline
    \end{tabular}
    \caption{Comparison of DES-CBC with Entropy Expansion for C++ rand(), /dev/urandom and /dev/random 1GB random numbers}
    \label{table3}
\end{table}
\end{center}
\begin{center}
\begin{table}[h!]
\centering
    \begin{tabular}{| m{2cm} | m{2cm} | m{2cm} | m{2cm} | m{2cm} | m{2cm} | m{2cm} | m{2cm} |}
        \hline
        Parameter & AES-256 on C++ rand()& Modified C++ rand()& AES-256 on /dev/urandom& Modified /dev/urandom& AES-256 on /dev/random& Modified /dev/random&Ideal Value\\
        \hline \hline
        Entropy & 8.000000 & 8.000000 & 8.000000 & 8.000000 & 8.000000 & 8.000000 &8.000000\\
        \hline
        Chi-Square Distribution & 234.06 & 271.33 & 247.86 & 254.60 & 295.12 & 239.96 &256.00 \\
        \hline
        Arithmetic Mean & 127.5005 & 127.5002 & 127.5041 & 127.4985 & 127.5037 & 127.5008 &127.5\\
        \hline
        Monte Carlo value of Pi & 3.141616169 & 3.141684931 & 3.141542811 & 3.141554084 & 3.141491916 & 3.141517494 &3.141592653\\
        \hline
        Serial Correlation Coefficient & 0.000000 & 0.000022 & -0.000017 & 0.000019 & 0.000043 & 0.000032 &0.0\\
        \hline
    \end{tabular}
    \caption{Comparison of AES-256-CTR with Entropy Expansion for C++ rand(), /dev/urandom and /dev/random 1GB random numbers}
    \label{table4}
\end{table}
\end{center}
\begin{center}
\begin{table}[h!]
\centering
    \begin{tabular}{| m{2cm} | m{2cm} | m{2cm} | m{2cm} | m{2cm} | m{2cm} | m{2cm} | m{2cm} |}
        \hline
        Parameter & QRNG File 1& Output File 1& QRNG File 2& Output File 2& QRNG File 3 & Output File 3 & Ideal Value\\
        \hline \hline
        Entropy & 7.999990 & 8.000000 & 8.000000 & 8.000000 & 8.000000 & 8.000000 &8.000000\\
        \hline
        Chi-Square Distribution & 265.55 & 236.57 & 282.64 & 255.05 & 266.87 & 269.28 & 256.00 \\
        \hline
        Arithmetic Mean & 127.4810 & 127.4994 & 127.5030 & 127.5014 & 127.4977 & 127.5003 & 127.5\\
        \hline
        Monte Carlo value of Pi & 3.142213394 & 3.141519305 & 3.141527487 & 3.141333663 &3.141651890 & 3.141697103 & 3.141592653\\
        \hline
        Serial Correlation Coefficient & -0.000530 & 0.000017 & 0.000014 & 0.000012 &-0.000031& 0.000008& 0.0\\
        \hline
    \end{tabular}
    \caption{Results for Tropos QRNG File-1 (raw data file), File-2 (10\% compression using Toeplitz matrix based PA) and File-3 (20\% compression). Note that all files contain around 1GB of random numbers.}
    \label{table5}
\end{table}
\end{center}
\begin{center}
\begin{table}[h!]
\centering
    \begin{tabular}{| m{2cm} | m{2cm} | m{2cm} | m{2cm} | m{2cm} | m{2cm} | m{2cm} | m{2cm} |}
        \hline
        Parameter & ID Quantique& Output File 1& Crypta Labs& Output File 2& Classical TRNG& Output File 3 & Ideal Value\\
        \hline \hline
        Entropy & 7.999999 & 7.999999 & 7.999998 & 7.999998 & 8.000000 & 8.000000 &8.000000\\
        \hline
        Chi-Square Distribution & 231.00 & 265.90 & 247.79 & 260.48 & 240.82 & 285.03 & 256.00 \\
        \hline
        Arithmetic Mean & 127.4973 & 127.4978 & 127.5075 & 127.4961 & 127.5050 & 127.5010 & 127.5\\
        \hline
        Monte Carlo value of Pi & 3.141995186 & 3.141344618 & 3.141686046 & 3.141344522 &3.141373392 & 3.141479930 & 3.141592653\\
        \hline
        Serial Correlation Coefficient & -0.000092 & 0.000036 & -0.000092 & 0.000073 &-0.000039& -0.000024& 0.0\\
        \hline
    \end{tabular}
    \caption{Results for ID Quantique's QRNG Data (125 MB) (File-1), Crypta Labs' QRNG Data (100 MB) (File-2) and Classical TRNG Data (1.3 GB) (File-3)}
    \label{table6}
\end{table}
\end{center}
\end{widetext}
\subsection{Comparison with Cryptographic Hash Function SHA-256}
For the 1GB random numbers generated above (using C++ rand(), /dev/urandom and /dev/random), we applied the SHA-256 hashing algorithm for every 160 bytes of these files and compared the randomness of the results with the entropy expansion whitening algorithm using the ENT test. Note that while using SHA-256 reduces the file size (in this case by a factor of 5 as 160 bits are used at a time and the output of the SHA-256 algorithm is 32 bytes or 256 bits long), entropy expansion does not reduce the file size. As entropy is determined by statistical testing, this is favourable for entropy of the resulting data as can be seen in the data compiled (Table \ref{table2}). For numbers generated by C++ rand(), all the parameters were observed to be better for the entropy expanded data than for the SHA-256 hashed numbers. Similarly, for numbers generated by /dev/urandom, entropy expansion showed better results than SHA-256 for all parameters. For numbers generated by /dev/random, it was seen that entropy expansion had better results for 3 parameters (entropy, arithmetic mean and Monte Carlo value of $\pi$) while SHA-256 had better results for chi-square distribution and the serial correlation coefficient.
\subsection{Comparison with Block Cipher DES-CBC}
The random numbers generated by the three PRNGs were also encrypted using the block cipher DES-CBC, which was the encryption standard until it was replaced by AES. Unlike SHA-256, block ciphers like DES or AES-256 do not reduce the size of files. Randomness comparison was done using both ENT and NIST. In the ENT tests, it was observed that for C++ rand() data, entropy expansion showed better results than DES for all parameters except the chi-square distribution (Table \ref{table3}). For /dev/urandom data, all parameters except the arithmetic mean showed better results for entropy expansion (Table \ref{table3}). For /dev/random data, all parameters showed better results for entropy expansion except the serial correlation coefficient (Table \ref{table3}). \\
For C++ rand() data and /dev/random data, it was seen that after encryption using DES-CBC, the files passed all the NIST tests (Refer \ref{randdes} for C++ rand() and \ref{randomdes} for /dev/random) which was also the case when entropy expansion was applied (Refer \ref{randmod} for C++ rand() and \ref{randommod} for /dev/random). For /dev/urandom however, it was seen that even after encryption with DES-CBC, the data failed for one NIST test (Refer \ref{urandomdes}, while when using entropy expansion it passed all the tests (Refer \ref{urandommod}).Note that entropy for all the files is the maximum possible so that cannot be a point of comparison here.\\
\subsection{Comparison with Block Cipher AES-256-CTR}
\begin{figure}[h!]
    \centering
    \resizebox{9.5cm}{8.5cm}{\includegraphics{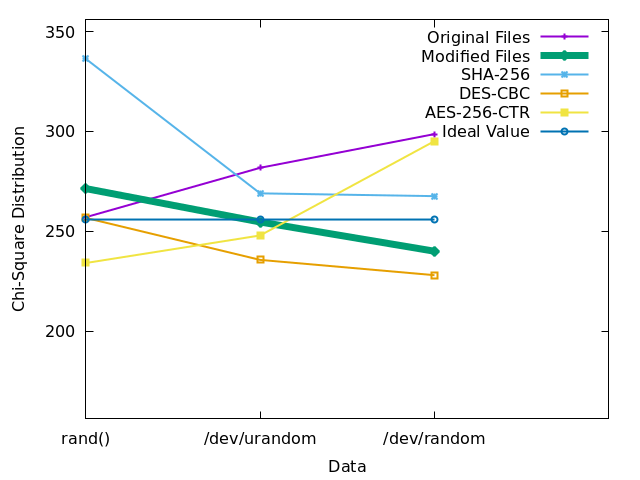}}
    \caption{Variation of Chi-Square Distribution of the Original Data and after Various Modifications}
    \label{chi}
\end{figure}

\begin{figure}[h!]
    \centering
    \resizebox{9.5cm}{8.5cm}{\includegraphics{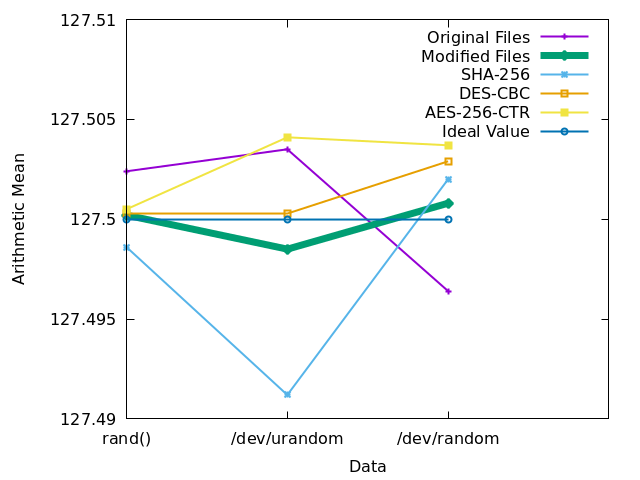}}
    \caption{Variation of Arithmetic Mean of the Original Data and after Various Modifications}
    \label{am}
\end{figure}

Next, the random numbers generated by the three PRNGs were also encrypted using the block cipher AES-256-CTR, which is one of the most commonly used encryption schemes. Again, ENT and NIST tests both were performed. For C++ rand() data, entropy expansion showed better results for chi-square distribution and arithmetic mean but AES-256-CTR showed better results for Monte Carlo value of $\pi$ and the serial correlation coefficient (Table \ref{table4}). For /dev/urandom data, it was seen that entropy expansion showed better results for all parameters except the serial correlation coefficient (Table \ref{table4}). For /dev/random data, it was observed that entropy expansion showed better results than AES-256-CTR did for all the parameters involved. Again, entropy for all the files is maximum here so it can't be used for comparison.\\
For the NIST tests, all the AES-256-CTR encrypted files passed all the NIST tests (Refer \ref{randaes} for C++ rand(), \ref{urandomaes} for /dev/urandom and \ref{randomaes} for /dev/random encrypted with AES-256). This is the same as for the entropy expanded versions of these files (Refer \ref{randmod} for C++ rand(), \ref{urandommod} for /dev/urandom and \ref{randommod} for /dev/random).
The variation of the chi-square distribution and the arithmetic mean for all of the above data was plotted (Fig. \ref{chi} and Fig. \ref{am}) and it was seen that the modified files after entropy expansion (bold green colour) gave some of the best results with respect to the ideal values of these quantities.

\subsection{Classical TRNG Data}
Lastly, we tested the effect of entropy expansion on some data (around 1.3 GB) generated by a non-quantum classical TRNG. The file was modified and the ENT test was performed. It was observed that except the chi-square distribution, all the randomness parameters were showing improvement (Table \ref{table6}).

\subsection{Quantum Random Number Generators (QRNGs - Tropos, ID Quantique and Crypta Labs)}

The whitening effect of the entropy expansion algorithm on quantum random number generators is observed. For this, we took three files with data (around 1GB each) generated using a QRNG and with their privacy amplified to different degrees were also taken and the ENT and NIST tests were performed on them. All the files were modified using entropy expansion using 32 matrices of size $8192 \times 8192$. \\
For file 1, it was observed that all the ENT parameters improved (Table \ref{table5}). The original file failed two NIST tests (Refer \ref{trunc}) while the modified file passed all of the NIST tests (Refer \ref{truncmod}).\\
For file 2, all the ENT parameters with the exception of the Monte Carlo value of $\pi$ showed an improvement after entropy expansion (Table \ref{table5}). In this case, it was seen that the original file performed poorly on the NIST test (Refer \ref{qrng2}) and showed a very low P-value for more than 40 tests. After entropy expansion, the randomness showed significant improvement and the modified file only failed 1 NIST test (Refer \ref{qrng2mod}).\\
For file 3, after entropy expansion the arithmetic mean and serial correlation coefficient improved while the chi-square distribution and Monte Carlo value of $\pi$ were better for the original file (Table \ref{table5}). In the NIST tests, the original data failed 2 tests (Refer \ref{qrng3}), but after application of entropy expansion it was observed that the modified data passed all the NIST tests (Refer \ref{qrng3mod}).\\
Next, we took 125 MB of data from ID Quantique's QRNG resource library \cite{IDquantique} and performed the ENT test on the file and its modified version using entropy expansion. It was seen that all parameters showed an improvement after entropy expansion (Table \ref{table6}).\\
100 MB of data was also taken from Crypta Labs \cite{crypt} and similarly modified using entropy expansion. The ENT test was performed and it was observed that all parameters except the Monte Carlo value of $\pi$ showed an improvement (Table \ref{table6}). 

\section{Conclusions}\label{Conclusion}
Random numbers are essential for scientific investigations and technological applications. High entropic random numbers are critical for cryptography.
Deterministic PRNGs and low entropic TRNGs create a
situation of entropy starvation, thus exposing data for cyber attacks. Recently, QRNGs have gained much attention because of their potential to harness ontic randomness from simple quantum phenomenon. However, manufacturing bias and technological imperfections give rise to imperfect randomness even in QRNGs. Hence, there is a requirement of post-processing to remove these biases. The existing methods on one hand require compression, thus decreasing the data rate and on other hand do not improve all the randomness parameters. 

In this work, we have addressed the problem
of universal whitening algorithm for the commercial random number generators by using n-qubit permutation matrices.
We have demonstrated the efficacy of our algorithm in several scenarios: PRNGs, classical TRNG as well as commerial QRNGs and also comparead it with popular cryptographic algorithms.
We have achieved significant improvement in the Chi-Square distribution value in majority (more than 70\%) of the instances. Note that other parameters also improve almost in every instance. The modified random data files after the application of our algorithm passes the NIST SP 800-22 tests in both the cases: 1. The raw file does not pass all the tests. 2. The raw file also passes all the tests.  
Our algorithm performs well in every scenario, thus has the potential to be a universal whitening algorithm for the commercial random number generators.

There are several offshoots of the present work. We
have demonstrated whitening mainly for 1 GB file size using 13-qubit permutation matrices (32 such matrices). One can determine the optimum matrix size and number of matrices for a given file size. The true potential of the algorithm will be shown with FPGA implementation and modification of the generated imperfect data at the runtime with the speed in Gbps. To ensure indeterminacy, one can use the raw data for the creation and selection of the permutation matrices.

\section*{Acknowledgement} 
S.G. acknowledges the Qunu Labs Pvt. Ltd. for the financial support.

\clearpage
\appendix
\onecolumngrid
\section{All NIST Test Suite Results}
\begin{lstlisting}[caption=NIST test results for 1GB data generated using C++ rand(), label={rand}]
------------------------------------------------------------------------------
RESULTS FOR THE UNIFORMITY OF P-VALUES AND THE PROPORTION OF PASSING SEQUENCES
------------------------------------------------------------------------------
   generator is <randcpp.bin>
------------------------------------------------------------------------------
 C1  C2  C3  C4  C5  C6  C7  C8  C9 C10  P-VALUE  PROPORTION  STATISTICAL TEST
------------------------------------------------------------------------------
115  85 134 107 117 105 120 127  93  97  0.020548   1091/1100    Frequency
132  97 113 100 111 104 112 114 110 107  0.582395   1088/1100    BlockFrequency
104 103 113 130 108 122 103  85 126 106  0.107628   1089/1100    CumulativeSums
110 107  95  91 125 129 130 112 103  98  0.063542   1088/1100    CumulativeSums
104 122 112 105 129 100 107 101 107 113  0.629548   1090/1100    Runs
125 117  85 113 117 101 117 107 113 105  0.328861   1090/1100    LongestRun
106 109  98 120 114  97 108 120 113 115  0.806575   1092/1100    Rank
130 114 110 109 110 119  98 107 110  93  0.480244   1085/1100    FFT
104 133 104 104 110 107 108 104 123 103  0.512317   1093/1100    NonOverlappingTemplate
123 117  95 116 104 109 107 110 123  96  0.525011   1091/1100    NonOverlappingTemplate
100 100 116  98 108 118 119 115 120 106  0.730786   1083/1100    NonOverlappingTemplate
112 133 110 105 122 111 108  89  96 114  0.193577   1088/1100    NonOverlappingTemplate
117 106 114  90 113 111 101 128 109 111  0.517744   1083/1100    NonOverlappingTemplate
117 112 111 115 112 127 109  94  99 104  0.621978   1091/1100    NonOverlappingTemplate
130 118 103  93 102  99 119  92 118 126  0.076572   1089/1100    NonOverlappingTemplate
107 110 106 119 124  84 105 102 118 125  0.195482   1088/1100    NonOverlappingTemplate
120 104 116  99 113  98 116 115 115 104  0.819817   1089/1100    NonOverlappingTemplate
106 109 113 111 102 104 124  99 124 108  0.754383   1091/1100    NonOverlappingTemplate
105 122 108 103 112 115 117  94  93 131  0.231300   1090/1100    NonOverlappingTemplate
114 123 110 103 119 103 112  97 101 118  0.719747   1088/1100    NonOverlappingTemplate
107 121  98 117 109 103 112 110 111 112  0.942579   1086/1100    NonOverlappingTemplate
 99 111 103 113  98 119 100 109 124 124  0.517744   1087/1100    NonOverlappingTemplate
101 108 102 142  97  99 117 100 109 125  0.053471   1087/1100    NonOverlappingTemplate
113 110 138 105 113 102 106 102 109 102  0.399601   1088/1100    NonOverlappingTemplate
106 130 115 113 109  99  97 112  98 121  0.420663   1089/1100    NonOverlappingTemplate
122 115  91  93  94 103 126 118 124 114  0.089449   1088/1100    NonOverlappingTemplate
115 116  92 117 114  98  97 120 125 106  0.347549   1084/1100    NonOverlappingTemplate
108 113 107  99 112 100 116 123 117 105  0.852898   1091/1100    NonOverlappingTemplate
 95 111 122 116 116 105 104 113 107 111  0.855909   1093/1100    NonOverlappingTemplate
109 105 139 101 110 113 104 108 104 107  0.394829   1086/1100    NonOverlappingTemplate
122 124 112 106  99 101 111 101  97 127  0.349015   1081/1100    NonOverlappingTemplate
115 113 100 115 109 101 128 108 116  95  0.580520   1090/1100    NonOverlappingTemplate
113 109 111 117 116  96 110 103 106 119  0.912605   1092/1100    NonOverlappingTemplate
114 103 107  88 123 111 111 111 111 121  0.559988   1086/1100    NonOverlappingTemplate
116 109 116 111  99 104 118  91 122 114  0.593666   1092/1100    NonOverlappingTemplate
102 107 105 134 113 123 113 108  94 101  0.280727   1089/1100    NonOverlappingTemplate
117  97 107 142  98 127 104  98 102 108  0.040587   1083/1100    NonOverlappingTemplate
 99 113 112 112  97  97 106 132 131 101  0.136570   1084/1100    NonOverlappingTemplate
118 103 103 110 102 134 115 121 100  94  0.221634   1089/1100    NonOverlappingTemplate
118 105 106 116 106 109 104 127 111  98  0.768644   1086/1100    NonOverlappingTemplate
105 115 117 102 112 113 108 101 107 120  0.948119   1089/1100    NonOverlappingTemplate
127  85 124 119  98 119 101 111  95 121  0.052549   1089/1100    NonOverlappingTemplate
114 110 119 105 113 121 113 109  90 106  0.723436   1086/1100    NonOverlappingTemplate
114 120 108 100 116 102 104 106 111 119  0.902838   1086/1100    NonOverlappingTemplate
110 108 120 113 120 105  97 102 108 117  0.854407   1090/1100    NonOverlappingTemplate
114  99 114 106 100 116 125 110 117  99  0.703044   1088/1100    NonOverlappingTemplate
104 112 119 117 103  99 105 113 102 126  0.708629   1094/1100    NonOverlappingTemplate
105 113  86 129 110 126  99  99 112 121  0.105903   1088/1100    NonOverlappingTemplate
117 113  97 109 117 106 109 103 113 116  0.939690   1087/1100    NonOverlappingTemplate
101 108 123 102 119 102 104 109 117 115  0.814889   1092/1100    NonOverlappingTemplate
122 102 118 111 115 107 107 102 102 114  0.899043   1088/1100    NonOverlappingTemplate
110 114 112 115 114 104 108 104 110 109  0.998591   1086/1100    NonOverlappingTemplate
111 125 103 127 104 107 108 108 108  99  0.663602   1090/1100    NonOverlappingTemplate
120 112 109 114 116 116 106 109  91 107  0.809915   1090/1100    NonOverlappingTemplate
115 117 108 116 104 105 100 103 121 111  0.907786   1084/1100    NonOverlappingTemplate
 97 123 116 109 110 111 119  92 124  99  0.351959   1091/1100    NonOverlappingTemplate
110 110  93 116 102 105 105 113 107 139  0.224821   1090/1100    NonOverlappingTemplate
111 134 122  96 113 102 100 108  98 116  0.237924   1087/1100    NonOverlappingTemplate
108 117  89 113 106 100 111 100 117 139  0.101421   1089/1100    NonOverlappingTemplate
114  92 115 118 123  98 119 127 102  92  0.135861   1085/1100    NonOverlappingTemplate
110 107 119 131 107 104 100 107 106 109  0.701179   1086/1100    NonOverlappingTemplate
113 111 103 112 111 127 113 102  95 113  0.739918   1092/1100    NonOverlappingTemplate
106 105 111  87 116 106 127 117 106 119  0.382270   1095/1100    NonOverlappingTemplate
104 103 125  99 121 102 121 106 106 113  0.629548   1091/1100    NonOverlappingTemplate
 95 113 105 133 127  96 126 108  98  99  0.060009   1092/1100    NonOverlappingTemplate
124 125 105 102  86 109 112 114 104 119  0.267086   1086/1100    NonOverlappingTemplate
 94 107 115 117 103  94  97 125 132 116  0.116622   1092/1100    NonOverlappingTemplate
108  99  97 106 127 122 100 119 111 111  0.492614   1091/1100    NonOverlappingTemplate
116 105  98 118  98 115 120 112 111 107  0.832720   1082/1100    NonOverlappingTemplate
125 103 104 114 116 108  93 117  90 130  0.134452   1088/1100    NonOverlappingTemplate
126 104 127 108 121 102 108  88 100 116  0.177983   1084/1100    NonOverlappingTemplate
111 112 128 108  87 100 112 113 112 117  0.406016   1090/1100    NonOverlappingTemplate
107 111 102 119 111 100 121 130  94 105  0.382270   1088/1100    NonOverlappingTemplate
122 117 127 110 101 100 120 112 108  83  0.135861   1092/1100    NonOverlappingTemplate
112 101 122 110 112 104 107  88 129 115  0.316327   1086/1100    NonOverlappingTemplate
119 114 105 100 116 108 101  95 117 125  0.569299   1087/1100    NonOverlappingTemplate
107  96  86 126 120 109 106  99 115 136  0.035598   1086/1100    NonOverlappingTemplate
106 106 104 109 119 115 125 110 102 104  0.871998   1093/1100    NonOverlappingTemplate
 93 114 114 112 123 100 114 118 113  99  0.604973   1091/1100    NonOverlappingTemplate
105 117 101 108 106 104 105 108 130 116  0.725277   1089/1100    NonOverlappingTemplate
111 106 100 105 124 108 103 120 114 109  0.866242   1090/1100    NonOverlappingTemplate
109 115  86 117 125 105 118 116 102 107  0.369957   1097/1100    NonOverlappingTemplate
126  96 110 107 109 102 120 109 117 104  0.691831   1084/1100    NonOverlappingTemplate
115 108 108 107 121 104  93 108 125 111  0.686204   1086/1100    NonOverlappingTemplate
124 134 100  89  95 110 131 103 110 104  0.027279   1084/1100    NonOverlappingTemplate
110 104 125 100 110  90 136 100 119 106  0.095010   1088/1100    NonOverlappingTemplate
109 121 105 118 122 109 111  98 100 107  0.784407   1086/1100    NonOverlappingTemplate
 93 127 121 124 105 106 101 115  94 114  0.216402   1092/1100    NonOverlappingTemplate
110 109 110  95 121 106 112  96 131 110  0.425610   1089/1100    NonOverlappingTemplate
 98 102 102 117 100 119 119 116 104 123  0.586147   1088/1100    NonOverlappingTemplate
 96 132 110 126 113 115 108  88 106 106  0.139438   1088/1100    NonOverlappingTemplate
 91 111 109 125 114 110 121 104 107 108  0.633333   1093/1100    NonOverlappingTemplate
117  97 105  94 126 110 108 118 113 112  0.574903   1088/1100    NonOverlappingTemplate
126 109 127 104 101 107  98 104 123 101  0.349015   1086/1100    NonOverlappingTemplate
103 115 114 112 114 120 121 102  91 108  0.646584   1088/1100    NonOverlappingTemplate
105 106 104  94 110 103 136 107 112 123  0.257477   1093/1100    NonOverlappingTemplate
125 117 104 117 122 102 124 105  90  94  0.156931   1085/1100    NonOverlappingTemplate
129 116  91  88 101 102 116 109 119 129  0.049288   1091/1100    NonOverlappingTemplate
103 109 122 101 115 114 100 121 128  87  0.179762   1091/1100    NonOverlappingTemplate
100 114 103 106 112 113  95 112 132 113  0.501531   1086/1100    NonOverlappingTemplate
 98 120 134 114  95 117 108 111  93 110  0.173596   1087/1100    NonOverlappingTemplate
123 110 109 107 114  97 115 115 109 101  0.874834   1079/1100    NonOverlappingTemplate
122 102 117 104 109 101 107 121 105 112  0.846799   1088/1100    NonOverlappingTemplate
102 109 117 127 108 122  94 103 101 117  0.440638   1086/1100    NonOverlappingTemplate
102 121 108 108 113  97 114 119 120  98  0.691831   1091/1100    NonOverlappingTemplate
112 114 132  93 105  97  92  99 136 120  0.017469   1084/1100    NonOverlappingTemplate
109 117 123  93 117  95 108 128 103 107  0.302791   1089/1100    NonOverlappingTemplate
122 109 107  83 108 125 115 111 106 114  0.301460   1086/1100    NonOverlappingTemplate
115 102 103 118 103 106 106 122 116 109  0.896474   1089/1100    NonOverlappingTemplate
100 107 119 115  91 114 106 112 111 125  0.573034   1092/1100    NonOverlappingTemplate
105 132 103 101 101  88 116 111 133 110  0.065007   1094/1100    NonOverlappingTemplate
126 130 107  94 104 112  94 105  98 130  0.062107   1091/1100    NonOverlappingTemplate
122 117 114 105  91  97 103 113  97 141  0.038244   1086/1100    NonOverlappingTemplate
111 116  98 122 113 108 112 111 115  94  0.772173   1085/1100    NonOverlappingTemplate
117 102 120  99 106 116 115 109 108 108  0.922967   1089/1100    NonOverlappingTemplate
 93 120 103 120 114 113 116 114 119  88  0.294867   1092/1100    NonOverlappingTemplate
109  97 118 104 107 119 116 117 106 107  0.892563   1086/1100    NonOverlappingTemplate
105 110 105  92 103 111 102 130 123 119  0.323249   1087/1100    NonOverlappingTemplate
113 116 101 110 102 132 112 102 106 106  0.633333   1092/1100    NonOverlappingTemplate
 97 122  93 108 107 124 114 102 121 112  0.415748   1090/1100    NonOverlappingTemplate
113 102 123 116 104 113 113 114  92 110  0.728952   1084/1100    NonOverlappingTemplate
112  95 104 114 114 106 101 105 118 131  0.494392   1086/1100    NonOverlappingTemplate
104 103 119  98 110 117 120 116 106 107  0.857405   1088/1100    NonOverlappingTemplate
108 111 102 108 123 109 106 104 116 113  0.960581   1090/1100    NonOverlappingTemplate
112 117 118 105 115 111 130  99 111  82  0.145326   1085/1100    NonOverlappingTemplate
116 115 115 114  99 118 114 100 120  89  0.476736   1089/1100    NonOverlappingTemplate
108 117  95 105 112 108 121 111 121 102  0.777440   1085/1100    NonOverlappingTemplate
 94 124 110 105 107 106 124 110 113 107  0.688081   1086/1100    NonOverlappingTemplate
115 111  97 107 105 103  97 108 131 126  0.316327   1092/1100    NonOverlappingTemplate
134 100  89 105 122 102 108 125 112 103  0.095530   1091/1100    NonOverlappingTemplate
111 106 114 111 113  97 116 109 117 106  0.969783   1090/1100    NonOverlappingTemplate
110 120 119 111 103  99 117 103 100 118  0.780932   1083/1100    NonOverlappingTemplate
122  98 117 115 111 106  86 118 105 122  0.289667   1087/1100    NonOverlappingTemplate
105 105 122  97 118 105 107 120 105 116  0.773933   1084/1100    NonOverlappingTemplate
113 102 109 120 112 114 105 117 106 102  0.957452   1088/1100    NonOverlappingTemplate
118 118 119 102 105 108 122 111 102  95  0.669264   1090/1100    NonOverlappingTemplate
127 109 118  90  94 113 118 109 114 108  0.347549   1085/1100    NonOverlappingTemplate
108  91 114 106 113 123  99 127 100 119  0.304126   1091/1100    NonOverlappingTemplate
 98 123 109 112  99 104 110 112 116 117  0.823075   1091/1100    NonOverlappingTemplate
105 105 116 109  94 104 136 113 108 110  0.374545   1091/1100    NonOverlappingTemplate
119  90 113 115 121 109  98 116 113 106  0.569299   1094/1100    NonOverlappingTemplate
138 111  98 109 108 108 102 108 114 104  0.382270   1090/1100    NonOverlappingTemplate
123 108 113 114 108 113  88 119 111 103  0.584270   1087/1100    NonOverlappingTemplate
103 102 113 108 111 102 107 114 130 110  0.779188   1091/1100    NonOverlappingTemplate
103 121 120 100 101 109  88 124 105 129  0.129620   1089/1100    NonOverlappingTemplate
105 119 128  93 110 109 123 108 102 103  0.423957   1082/1100    NonOverlappingTemplate
106 100 123 119 120 127 104 115 100  86  0.138716   1089/1100    NonOverlappingTemplate
120 109 113 121 111 103 119 110  93 101  0.673036   1086/1100    NonOverlappingTemplate
101 104 118 111 125 109 122  96 113 101  0.573034   1095/1100    NonOverlappingTemplate
108 110 115 118  97 101 114 119 120  98  0.717899   1089/1100    NonOverlappingTemplate
105 133 103 104 110 106 110 103 123 103  0.496173   1093/1100    NonOverlappingTemplate
111 118 115 112  94 120 107  97 119 107  0.686204   1088/1100    NonOverlappingTemplate
112 112 122 111 104 113 116  83 121 106  0.350485   1094/1100    NonOverlappingTemplate
 92 112 104 122 100 102 116 105 131 116  0.275709   1091/1100    NonOverlappingTemplate
118 115 121 107 107 115 107  97 109 104  0.893874   1090/1100    NonOverlappingTemplate
 97 113 114 109 111 111 103 128 104 110  0.787866   1091/1100    NonOverlappingTemplate
115 125 104  94 117 102 103 123 103 114  0.482003   1083/1100    OverlappingTemplate
126 111 102 104 118 114 110  98 113 104  0.770410   1090/1100    Universal
124 109 121 115 118 101 119  96  91 106  0.320468   1090/1100    ApproximateEntropy
 69  60  88  63  51  74  74  65  61  81  0.070017    677/686     RandomExcursions
 62  70  55  72  67  81  70  70  70  69  0.739918    678/686     RandomExcursions
 75  67  70  58  65  70  59  85  64  73  0.484646    677/686     RandomExcursions
 69  71  59  67  72  58  78  69  76  67  0.788728    680/686     RandomExcursions
 60  62  79  80  58  76  83  67  66  55  0.135446    681/686     RandomExcursions
 67  81  59  70  64  71  80  68  69  57  0.531185    677/686     RandomExcursions
 53  63  71  54  77  74  86  75  69  64  0.117189    680/686     RandomExcursions
 63  63  67  42  74  99  74  74  75  55  0.000509    674/686     RandomExcursions
 64  64  72  84  56  69  74  64  73  66  0.555020    679/686     RandomExcursionsVariant
 64  69  68  76  70  68  63  74  69  65  0.985918    679/686     RandomExcursionsVariant
 66  76  60  73  61  77  73  58  57  85  0.208381    678/686     RandomExcursionsVariant
 70  66  78  61  74  67  62  65  66  77  0.854318    679/686     RandomExcursionsVariant
 67  70  69  75  52  62  70  77  70  74  0.643244    680/686     RandomExcursionsVariant
 64  67  62  69  75  71  55  86  70  67  0.439994    678/686     RandomExcursionsVariant
 66  64  59  64  82  70  67  76  71  67  0.763217    680/686     RandomExcursionsVariant
 63  66  63  65  68  66  78  67  69  81  0.836865    684/686     RandomExcursionsVariant
 65  64  57  72  89  69  63  85  58  64  0.085434    684/686     RandomExcursionsVariant
 70  62  65  67  70  65  75  62  76  74  0.933071    680/686     RandomExcursionsVariant
 63  57  70  63  82  65  66  61  80  79  0.329484    679/686     RandomExcursionsVariant
 58  67  74  75  59  63  69  71  80  70  0.676859    680/686     RandomExcursionsVariant
 67  67  64  70  57  80  73  67  80  61  0.591287    679/686     RandomExcursionsVariant
 78  71  60  68  63  60  65  69  78  74  0.739918    680/686     RandomExcursionsVariant
 78  70  63  61  63  66  65  69  74  77  0.841931    681/686     RandomExcursionsVariant
 72  77  47  74  62  73  68  94  57  62  0.010280    679/686     RandomExcursionsVariant
 72  74  55  76  64  67  71  78  75  54  0.379806    679/686     RandomExcursionsVariant
 75  69  71  64  74  71  71  69  79  43  0.168002    682/686     RandomExcursionsVariant
118 109 116 104  97 109 135 102 106 104  0.390091   1088/1100    Serial
108 105 118 110 109 118 115 111  97 109  0.955013   1086/1100    Serial
 96 124 106  98  99 127  91 124 114 121  0.089449   1092/1100    LinearComplexity


- - - - - - - - - - - - - - - - - - - - - - - - - - - - - - - - - - - - - - - - -
The minimum pass rate for each statistical test with the exception of the
random excursion (variant) test is approximately = 1079 for a
sample size = 1100 binary sequences.

The minimum pass rate for the random excursion (variant) test
is approximately = 671 for a sample size = 686 binary sequences.

For further guidelines construct a probability table using the MAPLE program
provided in the addendum section of the documentation.
- - - - - - - - - - - - - - - - - - - - - - - - - - - - - - - - - - - - - - - - -
\end{lstlisting}
\begin{lstlisting}[caption=NIST test results for 1GB modified C++ rand() data , label={randmod}]
------------------------------------------------------------------------------
RESULTS FOR THE UNIFORMITY OF P-VALUES AND THE PROPORTION OF PASSING SEQUENCES
------------------------------------------------------------------------------
   generator is <randcpp_8192_32.bin>
------------------------------------------------------------------------------
 C1  C2  C3  C4  C5  C6  C7  C8  C9 C10  P-VALUE  PROPORTION  STATISTICAL TEST
------------------------------------------------------------------------------
114  93 114 112 130 103 109 127 104  94  0.195482   1090/1100    Frequency
 98 118 116  93 107 115 118 124 102 109  0.523191   1084/1100    BlockFrequency
105 110 113 119 102 118 121  97 111 104  0.818179   1089/1100    CumulativeSums
107 111  92  96 136 113 127 112 107  99  0.093976   1088/1100    CumulativeSums
125 109 106 100 114 112 111 104 117 102  0.863323   1088/1100    Runs
107 120 126 104 103 112 109 111 103 105  0.849861   1082/1100    LongestRun
104  93 128 114  97 118 106 109 129 102  0.203252   1092/1100    Rank
108 117 125 103 120  96 122 100  99 110  0.438954   1082/1100    FFT
113  99 100 104 112 108 125 124 105 110  0.684327   1085/1100    NonOverlappingTemplate
119 112  97 132 102  94 110  97 106 131  0.082788   1091/1100    NonOverlappingTemplate
101 128 120 117 104  90 114 105 103 118  0.319084   1094/1100    NonOverlappingTemplate
106  95 124 131  90  90 121 124 111 108  0.033122   1085/1100    NonOverlappingTemplate
 98 100 111 108  89 118 121 126 126 103  0.160154   1089/1100    NonOverlappingTemplate
 97 106 120 125 118 115 106  96 115 102  0.515932   1089/1100    NonOverlappingTemplate
107 100 108 116  87 124 119 101 128 110  0.193577   1086/1100    NonOverlappingTemplate
101 105 104 105 100 124 110 124 113 114  0.717899   1094/1100    NonOverlappingTemplate
 94 100 111 131 107 134 100 113 109 101  0.111747   1090/1100    NonOverlappingTemplate
 90  97 116 139  86 106 109 115 125 117  0.011567   1093/1100    NonOverlappingTemplate
128 102 101 112 101 120 110  86 120 120  0.162606   1086/1100    NonOverlappingTemplate
120 112 104 122 101 104  99 109 105 124  0.661713   1088/1100    NonOverlappingTemplate
104 122 120  92 105 119 109  95 120 114  0.386951   1089/1100    NonOverlappingTemplate
100 113 106 114 121 111  90 113 120 112  0.631440   1089/1100    NonOverlappingTemplate
112 105 113 126  83 117 103  97 119 125  0.105333   1090/1100    NonOverlappingTemplate
 98 100 109 126 106 112 107 112 115 115  0.806575   1093/1100    NonOverlappingTemplate
107 105 118 109 105 110 109 105 113 119  0.988255   1090/1100    NonOverlappingTemplate
100 104 100 116 101 126 109  99 125 120  0.383827   1093/1100    NonOverlappingTemplate
105  97 116 117 108 121 116 107 100 113  0.827923   1090/1100    NonOverlappingTemplate
119  89 106 117 108 109 127 112 114  99  0.410866   1092/1100    NonOverlappingTemplate
102 115 111 109 114  97 115 112 106 119  0.932645   1090/1100    NonOverlappingTemplate
107 104 107 107 103 130 122 115 106  99  0.591785   1089/1100    NonOverlappingTemplate
101 103 100 116 121 106 129 113  94 117  0.366918   1090/1100    NonOverlappingTemplate
113 106 100 118  98 115 113 113 109 115  0.932645   1089/1100    NonOverlappingTemplate
 96 111 108 113 114 102  96 108 132 120  0.369957   1089/1100    NonOverlappingTemplate
105  90  99 106 120 121 105 121 115 118  0.414117   1090/1100    NonOverlappingTemplate
104 109 115 116  97 120 119  96 112 112  0.747175   1087/1100    NonOverlappingTemplate
114 133 113 103 106  97 106 111 118  99  0.437274   1087/1100    NonOverlappingTemplate
101 123 102 119 114  94 110  99 114 124  0.428925   1089/1100    NonOverlappingTemplate
122 127 110 109 106 111  90 104 105 116  0.473239   1088/1100    NonOverlappingTemplate
105 106 124 123 112 107 104 101 109 109  0.843712   1088/1100    NonOverlappingTemplate
106 104 117 117 105 105  99 128 106 113  0.730786   1092/1100    NonOverlappingTemplate
114 109  94 103 130 109 114  87 122 118  0.144579   1088/1100    NonOverlappingTemplate
121 101 105 116  99 130 100 100 104 124  0.271989   1088/1100    NonOverlappingTemplate
101 100 141 105 113 117 102  96 118 107  0.116622   1088/1100    NonOverlappingTemplate
100 105 116 111 102 112 114 120  99 121  0.803215   1090/1100    NonOverlappingTemplate
113 122  91 112 112 117 107 103 119 104  0.659823   1084/1100    NonOverlappingTemplate
111 111 112 109 103 133 110  98 112 101  0.595549   1091/1100    NonOverlappingTemplate
105 105 120 115 101 109 112 100 122 111  0.867692   1094/1100    NonOverlappingTemplate
107 107 126 123 110 110  96 108 111 102  0.695575   1089/1100    NonOverlappingTemplate
100 114 111 112  87 123 123 115 101 114  0.343176   1096/1100    NonOverlappingTemplate
130 122 118 112 105  99 115 103  99  97  0.334538   1090/1100    NonOverlappingTemplate
111 108 103 122 113 113  95 103 125 107  0.680567   1083/1100    NonOverlappingTemplate
 97 109 118 129 120  92  98 105 115 117  0.244691   1085/1100    NonOverlappingTemplate
113 101 120 135  90 103 108 112  93 125  0.062107   1093/1100    NonOverlappingTemplate
122 129 119 104 109 104  97 100 112 104  0.455937   1084/1100    NonOverlappingTemplate
 95 126  91  96 113 114 100 118 102 145  0.006489   1090/1100    NonOverlappingTemplate
 96 116 121 106 114  96 100 108 118 125  0.450808   1092/1100    NonOverlappingTemplate
116 100 112 115 117 108  97 104 128 103  0.612525   1092/1100    NonOverlappingTemplate
103 107 103 128 113 112 113  93 124 104  0.450808   1088/1100    NonOverlappingTemplate
116 106 114 105 111 141  88 114 109  96  0.068416   1091/1100    NonOverlappingTemplate
102 117 115 119 106  92 109 138  99 103  0.130985   1086/1100    NonOverlappingTemplate
104  95 125 104 114 117 101 117 107 116  0.644691   1092/1100    NonOverlappingTemplate
113 108 116 107 107 101 111 109 109 119  0.989623   1092/1100    NonOverlappingTemplate
 88 138 112 108 109 103 128  95 117 102  0.034339   1090/1100    NonOverlappingTemplate
107 105 107 105 100 117 108 123 127 101  0.646584   1093/1100    NonOverlappingTemplate
106 128  92 115  96 112 122 110 110 109  0.401199   1090/1100    NonOverlappingTemplate
124 122 111 112  97  94 103 106 114 117  0.515932   1090/1100    NonOverlappingTemplate
104 104 119 109 117 111 118 107 106 105  0.968463   1087/1100    NonOverlappingTemplate
114 104 114 117 101 101 111 107 110 121  0.928429   1082/1100    NonOverlappingTemplate
120 112  92 105 109 128 102 100 120 112  0.391667   1083/1100    NonOverlappingTemplate
115 115 109 116 100 110 110 109  97 119  0.912605   1090/1100    NonOverlappingTemplate
132 115  97 113 111 114  98  97  97 126  0.157732   1088/1100    NonOverlappingTemplate
 96  92 127 129 123 101 112 102 109 109  0.139438   1090/1100    NonOverlappingTemplate
 90 100 111 102 130 104 113 103 126 121  0.160154   1090/1100    NonOverlappingTemplate
121 116 114 120 122 105 110  99  91 102  0.438954   1088/1100    NonOverlappingTemplate
101 107 118 108 109 109 102 111 124 111  0.921849   1091/1100    NonOverlappingTemplate
114 121 112 114 102 107  95 119 121  95  0.532315   1094/1100    NonOverlappingTemplate
 99 118 116 111 128 118  93  89 108 120  0.156931   1093/1100    NonOverlappingTemplate
106 106 109 123 109 121 113 111  99 103  0.869134   1093/1100    NonOverlappingTemplate
108 125  92 109  95 108 123 115 120 105  0.349015   1085/1100    NonOverlappingTemplate
114 101 124 112 118 102  98 103 111 117  0.732618   1091/1100    NonOverlappingTemplate
103 107  96 123 114 110 113 114 108 112  0.891244   1089/1100    NonOverlappingTemplate
106 104 122 104 112 119 122 113  97 101  0.684327   1089/1100    NonOverlappingTemplate
100 112 114 104 116 115 105 110 112 112  0.986385   1093/1100    NonOverlappingTemplate
132 112  91 121 108 126 103 102  93 112  0.094492   1083/1100    NonOverlappingTemplate
112 115  97 101 116 115 120 113 108 103  0.870570   1089/1100    NonOverlappingTemplate
120 116 115 129  99 106  95 106 111 103  0.471495   1094/1100    NonOverlappingTemplate
102  97 121 104 115 126 100 117 110 108  0.586147   1095/1100    NonOverlappingTemplate
113 102 112 112 123 112 112 108  88 118  0.599316   1091/1100    NonOverlappingTemplate
110  92 116 111 104 113 109 111 112 122  0.829528   1088/1100    NonOverlappingTemplate
102 116 114 121 112 105 117  97 111 105  0.864786   1089/1100    NonOverlappingTemplate
134 117  85 116  93 102 107 101 130 115  0.020676   1084/1100    NonOverlappingTemplate
120  95 109 117 105 111 107 102 120 114  0.801527   1082/1100    NonOverlappingTemplate
101 117 113 107  95 103 106 128 114 116  0.595549   1087/1100    NonOverlappingTemplate
 89 111 108 120 105 110 118 123 109 107  0.595549   1091/1100    NonOverlappingTemplate
106 116 114 104 112 106 103 102 108 129  0.791308   1092/1100    NonOverlappingTemplate
109  95 110 114 108 124  97 105 116 122  0.593666   1085/1100    NonOverlappingTemplate
111 115 115 108 116 115 117  85 107 111  0.608748   1088/1100    NonOverlappingTemplate
132 121 114 111  83 101 111 106 110 111  0.154550   1084/1100    NonOverlappingTemplate
127 125 107 129  96  92 105 104 100 115  0.107050   1085/1100    NonOverlappingTemplate
107  98 118  93 110 115 123 131 105 100  0.242419   1090/1100    NonOverlappingTemplate
106 107 128  99 130 109 108 109 104 100  0.419021   1089/1100    NonOverlappingTemplate
115 102 112 110 121 108 128 106 102  96  0.591785   1089/1100    NonOverlappingTemplate
108 128 101 109 113 111 109 112 115  94  0.697444   1082/1100    NonOverlappingTemplate
113 108 107 123 119 110 112 102 112  94  0.793022   1091/1100    NonOverlappingTemplate
110  98  93 107 111 112 115 113 121 120  0.701179   1090/1100    NonOverlappingTemplate
 99 131 109 107 107 110 117 110 109 101  0.691831   1093/1100    NonOverlappingTemplate
112 127  99 107  85 110 117 117 128  98  0.100327   1094/1100    NonOverlappingTemplate
123 109 124  93 106 103 116 101 115 110  0.550717   1088/1100    NonOverlappingTemplate
111 106  89 110 109 117 115 120 108 115  0.738097   1084/1100    NonOverlappingTemplate
111 115 101 108 105 109 111 122  94 124  0.671151   1085/1100    NonOverlappingTemplate
 94 111 112 113 107 115 106 116 111 115  0.942579   1092/1100    NonOverlappingTemplate
110 110 127  93 113 118 116  94 107 112  0.466281   1094/1100    NonOverlappingTemplate
124 114 105 104 102  97 105 119 118 112  0.721592   1088/1100    NonOverlappingTemplate
111 102 115 128 122 103 119 109  92  99  0.326047   1085/1100    NonOverlappingTemplate
129 107  96 119  92 115 127  94 111 110  0.121680   1085/1100    NonOverlappingTemplate
108 115 101 103 105 112 119 120 100 117  0.858895   1092/1100    NonOverlappingTemplate
106 107 120 101 106 126 108 112 103 111  0.845259   1094/1100    NonOverlappingTemplate
 94 118 111 118 110 114 116 108 109 102  0.867692   1090/1100    NonOverlappingTemplate
118  96  96 103 118 116 109 105 113 126  0.519557   1084/1100    NonOverlappingTemplate
106 113 108 107 100 127 107 111 103 118  0.834308   1091/1100    NonOverlappingTemplate
110 107 116 103 122  99 112  96 102 133  0.313587   1088/1100    NonOverlappingTemplate
115 106 111 103 114 101  96 114 130 110  0.608748   1084/1100    NonOverlappingTemplate
105 121 111 107 105 123 122 118  92  96  0.366918   1089/1100    NonOverlappingTemplate
102 116 123 111 107 106 119  97 106 113  0.818179   1085/1100    NonOverlappingTemplate
108 124 114 108 112 111 105 103 103 112  0.955835   1086/1100    NonOverlappingTemplate
115 115 104 109 105 105 119 107  98 123  0.842160   1093/1100    NonOverlappingTemplate
109 110 131 112 110 104 107 113  99 105  0.752586   1087/1100    NonOverlappingTemplate
116 113  99  91 106 128 118 106 121 102  0.327452   1090/1100    NonOverlappingTemplate
113 112 127  88  82 127 118 109  96 128  0.008213   1086/1100    NonOverlappingTemplate
113 123 117  93  88 129 108 106 106 117  0.156134   1087/1100    NonOverlappingTemplate
105 116 118 114  97  99  88 124 124 115  0.207229   1095/1100    NonOverlappingTemplate
125 106 106 102 133 111 104 110 111  92  0.274465   1086/1100    NonOverlappingTemplate
109 109 105 104 106 112 115 111 120 109  0.992825   1088/1100    NonOverlappingTemplate
115 121  93  91 120 120 112 114  97 117  0.261050   1085/1100    NonOverlappingTemplate
 95 118 134 111 112 106  93 103 102 126  0.121038   1088/1100    NonOverlappingTemplate
104 112 140 101  96 117 105  96 116 113  0.124936   1091/1100    NonOverlappingTemplate
113 114 120 115 121 101  94  88 113 121  0.268305   1090/1100    NonOverlappingTemplate
113  92 119  97  94 124 108 114 110 129  0.168450   1087/1100    NonOverlappingTemplate
110 119  99 113 108 110 117 110  99 115  0.928429   1086/1100    NonOverlappingTemplate
 92 122  96 116 113 108 118 128 103 104  0.290961   1087/1100    NonOverlappingTemplate
109 105  97 114  94 120 106 118 111 126  0.512317   1087/1100    NonOverlappingTemplate
112 105 111 106 125 107 111 104 109 110  0.968463   1092/1100    NonOverlappingTemplate
118 118 102 121 101  93 129  93 108 117  0.181557   1091/1100    NonOverlappingTemplate
 96 121  94 112 131 111  99 114 107 115  0.288378   1094/1100    NonOverlappingTemplate
105 111  98 110 105 113 130 113 107 108  0.770410   1092/1100    NonOverlappingTemplate
122 111 110 108 107 101 125 118  99  99  0.637119   1092/1100    NonOverlappingTemplate
112 133 112 125 105 113 112  95 100  93  0.169299   1083/1100    NonOverlappingTemplate
 99 120 116 105 114 136  94  89 112 115  0.079183   1081/1100    NonOverlappingTemplate
115 115 108 103 103 104 106 107 117 122  0.930554   1084/1100    NonOverlappingTemplate
107  94 118 102 108 119 104 105 109 134  0.338839   1088/1100    NonOverlappingTemplate
113  99 100 104 111 109 125 124 105 110  0.689957   1085/1100    NonOverlappingTemplate
132 128  98 100 101 108  94 121  96 122  0.051043   1087/1100    NonOverlappingTemplate
113 113 105 124  98 115 105 117 112  98  0.766873   1089/1100    NonOverlappingTemplate
105 109 111 115 104 125  94 119  95 123  0.393246   1091/1100    NonOverlappingTemplate
 95 106 109  88 121  88 135 128 106 124  0.007021   1090/1100    NonOverlappingTemplate
115  94 111 103 117 101 117 129 123  90  0.158536   1085/1100    NonOverlappingTemplate
115 115 113 108  94  99 104 123 121 108  0.637119   1087/1100    OverlappingTemplate
123  91 119 108 123 103 102 115 114 102  0.427266   1088/1100    Universal
120 104  88 100  95 112 145 110 109 117  0.016613   1093/1100    ApproximateEntropy
 67  75  75  68  75  69  63  58  75  57  0.685984    675/682     RandomExcursions
 63  69  66  85  77  67  67  64  68  56  0.507706    676/682     RandomExcursions
 58  84  72  74  78  71  49  77  57  62  0.058765    676/682     RandomExcursions
 60  71  65  78  59  78  51  64  77  79  0.172146    677/682     RandomExcursions
 84  69  83  75  65  66  71  61  63  45  0.048716    670/682     RandomExcursions
 76  81  57  81  72  53  65  67  55  75  0.102885    673/682     RandomExcursions
 62  66  71  71  52  83  59  70  75  73  0.329484    677/682     RandomExcursions
 67  48  81  74  69  63  66  74  67  73  0.329484    674/682     RandomExcursions
 64  69  75  65  68  63  66  75  53  84  0.407961    679/682     RandomExcursionsVariant
 68  56  80  71  77  58  69  52  81  70  0.147275    680/682     RandomExcursionsVariant
 64  60  77  57  66  92  69  65  71  61  0.134308    678/682     RandomExcursionsVariant
 67  62  61  66  88  67  68  72  65  66  0.570068    678/682     RandomExcursionsVariant
 64  74  65  62  80  70  70  73  62  62  0.826550    678/682     RandomExcursionsVariant
 68  79  66  77  62  64  66  49  69  82  0.210013    675/682     RandomExcursionsVariant
 77  81  73  65  57  70  55  63  71  70  0.431862    674/682     RandomExcursionsVariant
 81  76  67  65  67  70  65  64  71  56  0.713161    676/682     RandomExcursionsVariant
 77  81  62  67  65  54  72  62  72  70  0.504800    674/682     RandomExcursionsVariant
 67  72  66  60  70  77  84  58  64  64  0.513539    674/682     RandomExcursionsVariant
 46  76  74  75  76  63  75  54  83  60  0.030090    678/682     RandomExcursionsVariant
 61  54  84  63  79  80  72  67  59  63  0.143638    677/682     RandomExcursionsVariant
 56  83  67  62  76  72  71  64  76  55  0.279763    675/682     RandomExcursionsVariant
 58  77  76  83  74  60  67  65  59  63  0.343389    677/682     RandomExcursionsVariant
 66  71  75  76  61  75  72  62  78  46  0.179237    676/682     RandomExcursionsVariant
 68  79  49  77  69  62  76  67  82  53  0.067498    678/682     RandomExcursionsVariant
 66  71  71  63  76  68  73  64  66  64  0.980082    677/682     RandomExcursionsVariant
 64  76  66  73  59  69  69  66  59  81  0.682945    678/682     RandomExcursionsVariant
115 112 103 118 102 110 107 106 107 120  0.960581   1084/1100    Serial
111 106 111  98 114 119 104 110 112 115  0.966418   1088/1100    Serial
117 122  95 109 106  99 121 112 114 105  0.682447   1090/1100    LinearComplexity


- - - - - - - - - - - - - - - - - - - - - - - - - - - - - - - - - - - - - - - - -
The minimum pass rate for each statistical test with the exception of the
random excursion (variant) test is approximately = 1079 for a
sample size = 1100 binary sequences.

The minimum pass rate for the random excursion (variant) test
is approximately = 667 for a sample size = 682 binary sequences.

For further guidelines construct a probability table using the MAPLE program
provided in the addendum section of the documentation.
- - - - - - - - - - - - - - - - - - - - - - - - - - - - - - - - - - - - - - - - -
\end{lstlisting}
\lstinputlisting[caption=NIST test results for 1GB data generated using /dev/urandom, label={urandom}]{urandom_nist.txt}
\lstinputlisting[caption=NIST test results for 1GB modified /dev/urandom data, label={urandommod}]{urandom_8192_32_report.txt}
\lstinputlisting[caption=NIST test results for 1GB data generated using /dev/random, label={random}]{random_report.txt}
\lstinputlisting[caption=NIST test results for 1GB modified /dev/random data, label={randommod}]{random_8192_32_report.txt}
\lstinputlisting[caption=NIST test results for 1GB C++ rand() data encrypted using DES-CBC, label={randdes}]{randcpp_des_report.txt}
\lstinputlisting[caption=NIST test results for 1GB /dev/urandom data encrypted using DES-CBC, label={urandomdes}]{urandom_des_report.txt}
\lstinputlisting[caption=NIST test results for 1GB /dev/random data encrypted using DES-CBC, label={randomdes}]{random_des_report.txt}
\lstinputlisting[caption=NIST test results for 1GB C++ rand() data encrypted using AES-256, label={randaes}]{randcpp_aes_report.txt}
\lstinputlisting[caption=NIST test results for 1GB /dev/urandom data encrypted using AES-256, label={urandomaes}]{urandom_aes_report.txt}
\lstinputlisting[caption=NIST test results for 1GB /dev/random data encrypted using AES-256, label={randomaes}]{random_aes_report.txt}
\lstinputlisting[caption=NIST test results for QRNG data (File 1), label={trunc}]{trunc.txt}
\lstinputlisting[caption=NIST test results for QRNG data (File 1) modified using entropy expansion, label={truncmod}]{truncmod.txt}
\lstinputlisting[caption=NIST test results for QRNG data (File 2), label={qrng2}]{QRNG2.txt}
\lstinputlisting[caption=NIST test results for QRNG data (File 2) modified using entropy expansion, label={qrng2mod}]{QRNG2mod.txt}
\lstinputlisting[caption=NIST test results for QRNG data (File 3), label={qrng3}]{QRNG3.txt}
\lstinputlisting[caption=NIST test results for QRNG data (File 3) modified using entropy expansion, label={qrng3mod}]{QRNG3mod.txt}
\end{document}